\documentstyle[twocolumn,eqsecnum,aps,epsfig]{revtex}
\newcommand{\be}{\begin{equation}}
\newcommand{\ee}{\end{equation}}
\newcommand{\bea}{\begin{eqnarray}}
\newcommand{\eea}{\end{eqnarray}}

\begin{document}
\title{Nonlinear evolution of the momentum dependent condensates in strong
interaction: the ``pseudoscalar laser''}
\author{Adrian Dumitru$^a$ and Ove Scavenius$^b$}
\address{$^a$ Physics Department, Columbia University,
538W 120th Street, New York, NY 10027, USA\\
$^b$ The Niels Bohr Institute, Blegdamsvej 17, DK-2100 Copenhagen {\O},
Denmark}
\date{\today}
\maketitle   
\begin{abstract}
We discuss the relaxation of the scalar and pseudoscalar condensates
after a rapid quench from an initial state with fluctuations. If we include
not only the zero-mode but also higher modes of the condensates in the
classical evolution, we observe parametric amplification of those
``hard'' modes. Thus, they couple nonlinearly
to the ``soft'' modes. As a consequence, domains of coherent
$\pi$-field emerge long after the initial spinodal decomposition.
The momentum-space distribution of pions emerging from the decay of that
momentum-dependent condensate is discussed.
\end{abstract}
\narrowtext

The exciting speculation that the dynamics of spontaneous breaking of
chiral symmetry could reorient the condensates in QCD
as compared to the physical vacuum has stimulated many
studies, see
e.g.~\cite{baked_alaska,pi_laser,RW,DomSize,Boyanovsky:1995yk,Scavenius:1999zc,Asakawa:1995wk},
and~\cite{BlaizotKrzywicki96} for reviews.
However, for the chiral symmetry breaking process in strong interaction
the spinodal decomposition after a rapid ``quench'' does not yield a large
correlation length~\cite{Boyanovsky:1993pf}.
Accordingly, a simulation~\cite{DomSize} of the classical equations of motion
of the linear $\sigma$-model
on a 3d-lattice with spacing $a=1$~fm confirmed that in the
strong-coupling case no large domains
form in which the $\pi$-field were essentially uniform, see
also~\cite{Boyanovsky:1995yk}.

We shall confirm that result but also argue that if the cutoff for
the Fourier modes of the classical order parameter of chiral symmetry
is pushed to larger values by some physical mechanism, e.g.\
parametric
resonance~\cite{Kofman:1994rk,Hiro-Oka:2000xk,Kaiser:1999hf},
domains much bigger than the Compton wavelength
of the pion {\em can} form over time-scales $\gg1$~fm.
Thus, the ``pseudoscalar laser'' is a non-linear
phenomenon developing long after spinodal decomposition.

Our rather simple idea is that
parametric resonance leads to large occupation numbers of ``hard''
modes of the scalar and pseudoscalar condensates, which then couple
non-linearly to the ``soft'' modes. Thus,
the same mechanism that is responsible for the
explosive heating of the universe after inflation~\cite{Kofman:1994rk}
(see also~\cite{Khlebnikov:1996mc} were the inflaton decay was studied
via a simulation on the lattice, similar to our studies here)
leads to the formation of domains where the pseudoscalar condensate
exhibits coherent oscillations with big amplitude around its vacuum
value.
Those domains could be produced in the laboratory in collisions of
protons~\cite{baked_alaska} or heavier nuclei~\cite{pi_laser}
at high energies.

As an effective theory of the chiral symmetry breaking
dynamics~\cite{Pisarski:1984ms}
we apply the linear $\sigma$-model~\cite{Gell-Mann:1960np}
\bea
{\cal L} &=&
 \overline{q}[i\gamma ^{\mu}\partial _{\mu}-g(\sigma +i\gamma _{5}
 \vec{\tau} \cdot \vec{\pi} )]q\nonumber\\
&+& \frac{1}{2}(\partial _{\mu}\sigma \partial ^{\mu}\sigma + \partial _{\mu}
\vec{\pi} \partial ^{\mu}\vec{\pi} )
- U(\sigma ,\vec{\pi})\quad.
\label{sigma}
\eea
The potential exhibiting the spontaneously broken symmetry is
\begin{equation} \label{potential}
U(\sigma ,\vec{\pi} )=\frac{\lambda ^{2}}{4}(\sigma ^{2}+\vec{\pi} ^{2} -
{\it v}^{2})^{2}-H\sigma.
\end{equation}
Here $q$ is the constituent-quark field $q=(u,d)$. The
scalar field $\sigma$ and the pseudoscalar field $\vec{\pi} 
=(\pi_{1},\pi_{2},\pi_{3})$ 
together form a chiral field $\Phi =(\sigma,\vec{\pi})$. 
The parameters of the 
Lagrangian are usually chosen such that the chiral $SU_{L}(2) \otimes
SU_{R}(2)$ symmetry is spontaneously 
broken in the vacuum and the expectation values of the condensates are 
$\langle\sigma\rangle ={\it f}_{\pi}$ and $\langle\vec{\pi}\rangle =0$, where
${\it f}_{\pi}=93$~MeV is the pion decay constant. The explicit
symmetry breaking term is determined
by the PCAC relation which gives $H=f_{\pi}m_{\pi}^{2}$, where
$m_{\pi}=138$~MeV is the pion mass. Then one finds $v^{2}=f^{2}_{\pi}-
{m^{2}_{\pi}}/{\lambda ^{2}}$. The $\sigma$-mass, $m^2_\sigma=2
\lambda^{2}f^{2}_{\pi}+m^{2}_{\pi}$, which we set to 600~MeV, yields 
$\lambda^{2}\approx20$. 
For the moment we focus exclusively on the dynamics of the chiral condensates,
i.e.\ we set $g=0$. We return to the case $g\ne0$ at the end.

We shall discuss the evolution of the system in the
potential~(\ref{potential}), starting from a chirally
symmetric initial state, $\langle \Phi\rangle=0$. Such an initial state
could be reached thanks to the slow dynamics of the long-wavelength
modes~\cite{RW,DomSize,Boyanovsky:1995yk,Boyanovsky:1993pf};
or, in the case $g>0$, due to
a possible first-order symmetry breaking phase transition which``traps'' the
expectation value of the fields in the local minimum around
$\langle \Phi\rangle=0$~\cite{Scavenius:1999zc}.
We include fluctuations in the initial state by distributing the
fields as Gaussian random variables with~\cite{RW,DomSize}
\be \label{inifluct}
\langle \Phi_a^2\rangle = v^2/16\quad,\quad
\langle \dot\Phi_a^2\rangle = v^2/4~{\rm fm}^2\quad.
\ee
Since $\langle \Phi\rangle=0$, the system is in an unstable
condition for the potential at hand, eq.~(\ref{potential}).
The fields ``roll down'' towards the vacuum. In analogy to condensed matter
physics this process is called spinodal
decomposition~\cite{DomSize,Boyanovsky:1993pf}.

\begin{figure}[htp]
\centerline{\hbox{\epsfig{figure=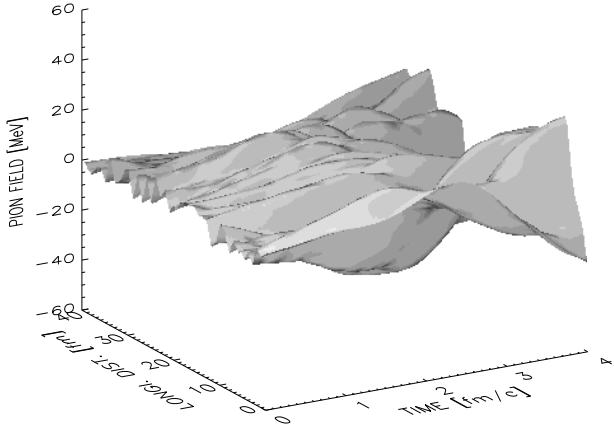,width=10cm}}}
\caption{Time evolution of the $\pi$-field in the longitudinal space-direction,
and isospin-0 direction,
averaged over the transverse directions, time $0-4$~fm (practically
independent of lattice spacing as long as $0.1$~fm$\le a\le1$~fm).}
\label{avfield_t20}
\end{figure}  
We first show numerical solutions of the classical equations of motion
obtained from~(\ref{sigma}) on a lattice of physical size 
10~fm$\times$10~fm in ``transverse'' directions, and 40~fm in longitudinal
direction. (Throughout the manuscript Minkowski-metric has been employed).
We have integrated the equations using a second-order leap-frog
algorithm, ensuring that energy and momentum were conserved to within $0.1\%$.
We performed simulations
with lattice spacing $a=1$~fm as well as $a=0.2$~fm. In the latter case,
a coarse-graining procedure was applied to the initial configuration to
obtain a correlation length of 1~fm; this was done without artificially cooling
the fluctuations~(\ref{inifluct}).

Fig.~\ref{avfield_t20} depicts the $\pi$-field averaged over the
transverse plane, showing the evolution within the first 4~fm.
One observes that the field amplitude within each individual small
initial ``domain'' gets amplified, exactly as it should be~\cite{RW}.
However, the wavelengths of the $\pi$-field oscillations
more or less correspond to those of a free pion.
Even for the rather small initial fluctuations~(\ref{inifluct})
one can hardly observe big domains emerging during the
initial ``roll-down'' \cite{Boyanovsky:1993pf,Boyanovsky:1995yk},
the less the larger the fluctuations at $t=0$.
Also, we have found that
the amplitude of the $\pi$-field after ``roll-down'' are almost
independent of $a$. This is no longer true for the long-time evolution
to be discussed next.

\begin{figure}[htp]
\centerline{\hbox{\epsfig{figure=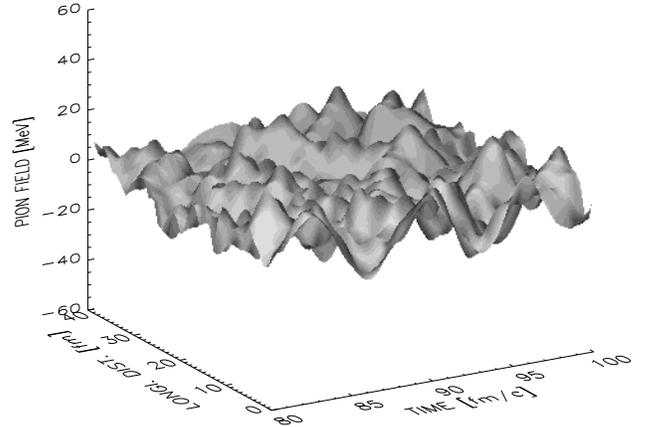,width=10cm}}}
\caption{As Fig.~\ref{avfield_t20} but at time $80-100$~fm (lattice
spacing $a=1$~fm).}
\label{avfield_t80a1}
\end{figure}

For large lattice spacing no big domains can be observed even at much
later times, $t=80-100$~fm, as shown in Fig.~\ref{avfield_t80a1}.
The field is basically exhibiting only small and
short-range oscillations. 

\begin{figure}[htp]
\centerline{\hbox{\epsfig{figure=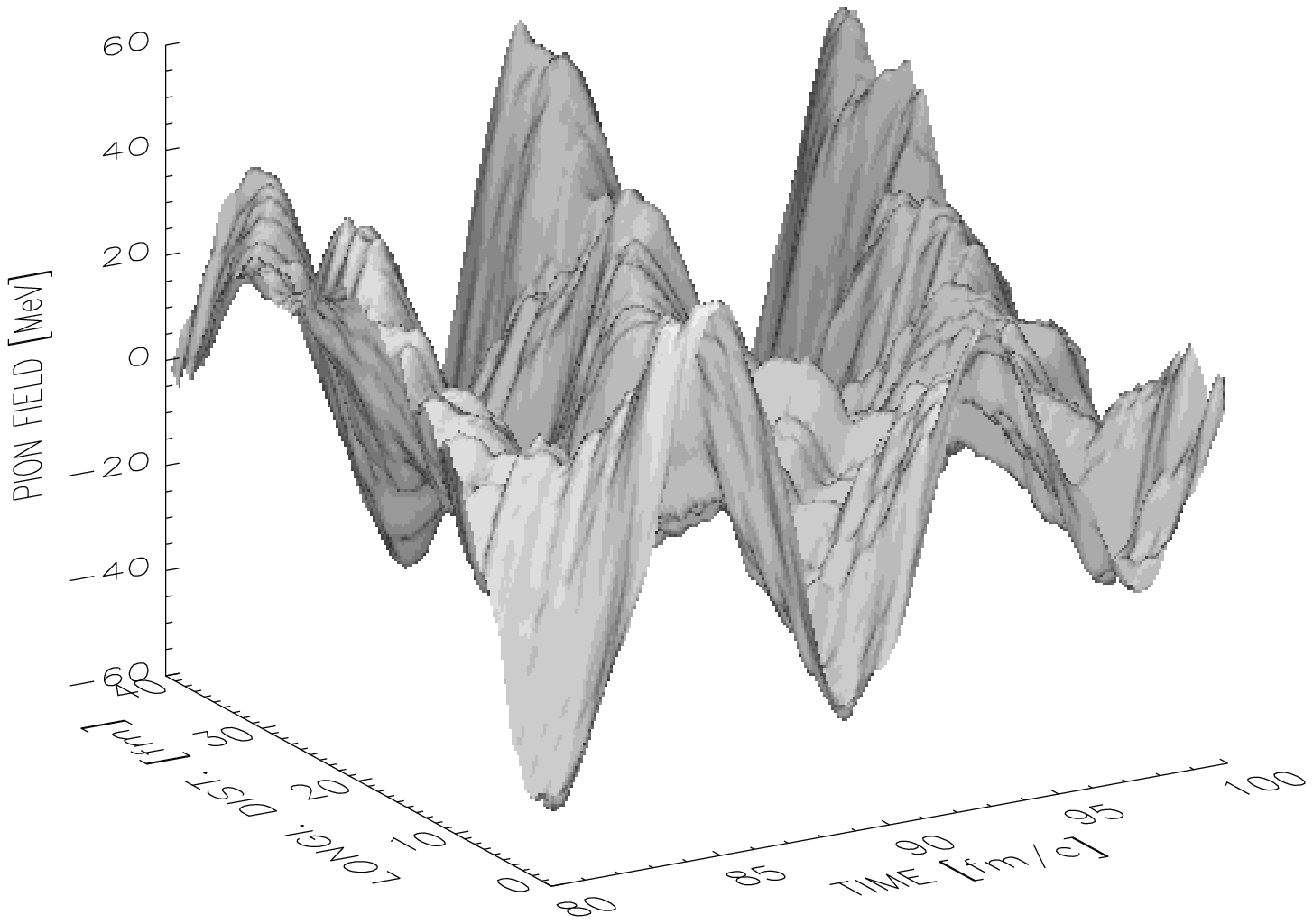,width=10cm}}}
\caption{As Fig.~\ref{avfield_t20} but at time $80-100$~fm (lattice
spacing $a=0.2$~fm).}
\label{avfield_t80}
\end{figure}

However, for $a=0.2$~fm, one can observe
long-range oscillations with a much bigger amplitude,
clearly exhibiting the strong amplification of the long-wavelength
modes and, correspondingly, big domains (Fig.~\ref{avfield_t80}).
Moreover, this observation remains true even for
larger initial fluctuations.

\begin{figure}[htp]
\centerline{\hbox{\epsfig{figure=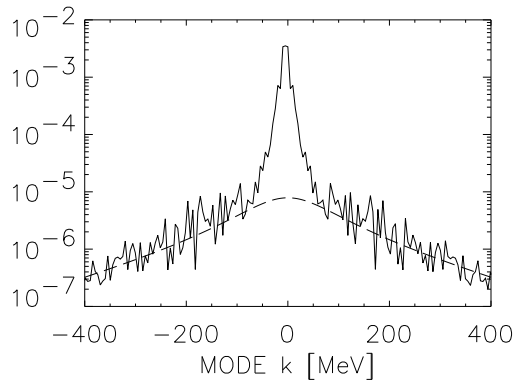,width=8cm}}}
\caption{Distribution of pions of given isospin 0 in momentum space,
$dN/d^3k$ (lattice spacing $a=0.2$~fm, time $t=100$~fm). The broken line
depicts a statistical (Bose) distribution.}
\label{pispec}
\end{figure}

In Fig.~\ref{pispec} we show the momentum-space distribution of
$\pi$-particles in the condensate
(the distribution observed if the field
decayed into particles at that moment) obtained from the spacial Fourier
transform of $\pi(t,\vec{x})$. 
For simplicity, when computing $dN/d^3k$ we assumed the Hartree dispersion
relation $\Omega^2=k^2+\lambda^2\left(\langle{\sigma^2}\rangle
+5\langle{{\pi}_0^2}\rangle/3-v^2\right)$.
(The even more simple-minded application of the free-pion dispersion
relation gave practically the same result.)
The result has been averaged over several initial field configurations,
chosen randomly according to~(\ref{inifluct}).
One observes a very sharp rise below $k\sim70$~MeV, proving the
spacial homogeneity of the pseudoscalar condensate at $t=100$~fm.
Note that a statistical (Bose-Einstein) distribution does not
show such a ``rise'', i.e.\ local maximum of the second derivative,
at non-vanishing $k$.
(In
Fig.~\ref{pispec}, the statistical distribution is for a temperature
$T=160$~MeV and a chemical potential $\mu_\pi=100$~MeV.)

To understand the results of the simulation we analyze the effect of
the $\sigma$-field oscillations on the Fourier modes of the
$\pi$-field in a more schematic model.
We employ the linearized classical equation of
motion for the pion field, neglecting also explicit symmetry breaking
($m_{\pi}=0$). Furthermore, we assume that the sigma-field makes
harmonic oscillations with amplitude $\sigma_0$ in the minimum of
the potential, $\sigma=\sigma_0\sin(m_\sigma t)$.
After Fourier transformation, the evolution of the
$\pi$-field modes becomes (see also~\cite{Kofman:1994rk})
\be
\ddot\pi_k+\left\{k^2+\lambda^2\sigma\left(
\sigma+2v\right)\right\}\pi_k=0\quad.
\ee
The expression in curly brackets gives the frequency $\Omega_k^2$ of the
$\pi$-field oscillations.
As initial condition for the $\pi$-field we assume
$\pi_k(t=0)=1/\sqrt{2\Omega_k}$, $\dot\pi_k(t=0)=\sqrt{\Omega_k/2}$.
The solution of this equation can be found numerically
without great difficulties (we used the routine ``odeint''
from~\cite{numrec}, modified to work with double precision,
and required accuracy $10^{-9}$).

\begin{figure}[htp]
\centerline{\hbox{\epsfig{figure=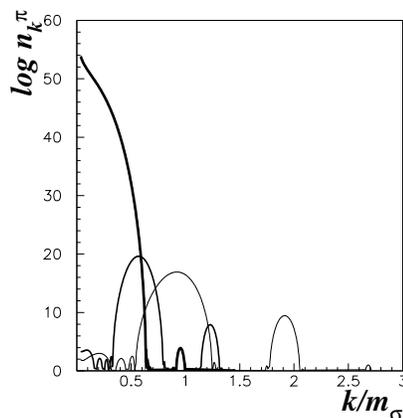,width=6cm}}}
\caption{Occupation numbers of various Fourier modes of the $\pi$-field
after 10 oscillations of the $\sigma$-field with amplitude $\sigma_0=v$
(thick line with very strong soft modes), $\sigma_0=2v$
(thinner line), and $\sigma_0=m_\sigma$ (thinnest line), respectively.}
\label{occno_b}
\end{figure}

Fig.~\ref{occno_b} shows the occupation number
\be
n_k^\pi = \frac{\Omega_k}{2}\left( \frac{\left|\dot\pi_k\right|^2}{\Omega_k^2}+
\left|\pi_k\right|^2\right)-\frac{1}{2}
\ee
of the $\pi$-field modes for three different $\sigma$-field
amplitudes.
(For
$\sigma_0\ll v$ we found much smaller occupation numbers and ``squeezed''
resonance bands.)
Oscillations with amplitude $\sigma_0\ge v$ will
certainly not be harmonic, see e.g.~\cite{Kaiser:1999hf}.
However, this is only a schematic and qualitative
example anyway, since it completely neglects the nonlinearities in the
$\pi$-field and the back-reaction on the $\sigma$,
which are most essential in the
lattice simulation~\cite{Khlebnikov:1996mc}.
We thus do not attempt to be more ``realistic'' in this respect.

The strong amplification of the soft
modes of the $\pi$-field, $k\sim0$ has been
interpreted as being due to an imaginary effective mass of the
pion~\cite{Asakawa:1995wk,BlaizotKrzywicki96}. 
However, there are additional resonance bands
at higher $k$, being
particularly pronounced if the amplitude of the $\sigma$-field is
big.

\begin{figure}[htp]
\centerline{\hbox{\epsfig{figure=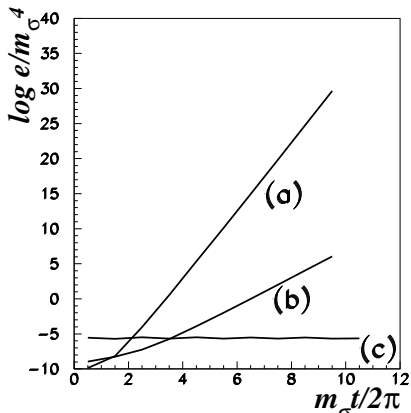,width=6cm}}}
\caption{Energy density of the $\pi$-field (exponentially increasing)
and of the quarks (constant; only modes up to $k=6m_\sigma$),
averaged over one period of oscillation of the $\sigma$-field.
$\sigma_0=v$ for curves (a,c), $\sigma_0=2v$ for curve (b).}
\label{edens}
\end{figure}

In Fig.~\ref{edens} we show the energy density of the $\pi$-field,
\be \label{edenseq}
e^\pi = 3\int\frac{d^3k}{(2\pi)^3} \Omega_k n_k^\pi\quad.
\ee
The factor 3 in front of the integral comes from the three-fold isospin
degeneracy of the $\pi$-field. This is the energy density that would
be measured if the classical field decayed into particles with
energy $\Omega_k$ at the corresponding time.
One observes that the energy density of the
$\pi$-field increases exponentially in time. In a more realistic
calculation, as e.g.\ the lattice simulation described above,
that exponential growth will shut off eventually due to
energy conservation and the back-reaction.

In a case like this we do not integrate out the $k>0$ modes, or the
modes above some cutoff $k_c\ll T$, and treat only the zero-mode
classically~\cite{Bodeker:1995pp} (evolving in the effective potential
generated by the ``hard'' modes above $k_c$).
In particular, we can not compute the dissipative
corrections to the classical dynamics of the soft modes, which is
due to scattering with the ``hard'' modes,
perturbartively (for an alternative approach see~\cite{Cooper:1997ii}).
We therefore decided to describe
all modes within the first few parametric resonance bands
classically and to solve the full nonlinear field equations
(on the lattice).

Of course, we can not include arbitrarily hard modes in our classical
description as we would run into the well-known Rayleigh-Jeans type
of ultraviolet divergences, cf.\ e.g.\ the discussion in~\cite{Bodeker:1995pp}.
Fortunately, our lattice simulations were practically independent of the
lattice spacing (domains always formed) as long as $a\simeq0.1-0.3$~fm,
showing that higher resonance bands do not contribute to the classical
evolution. One may have to employ smaller $a$ (i.e.\ larger $k_c$) if
the initial fluctuations were substantially larger than~(\ref{inifluct}).

\begin{figure}[htp]
\centerline{\hbox{\epsfig{figure=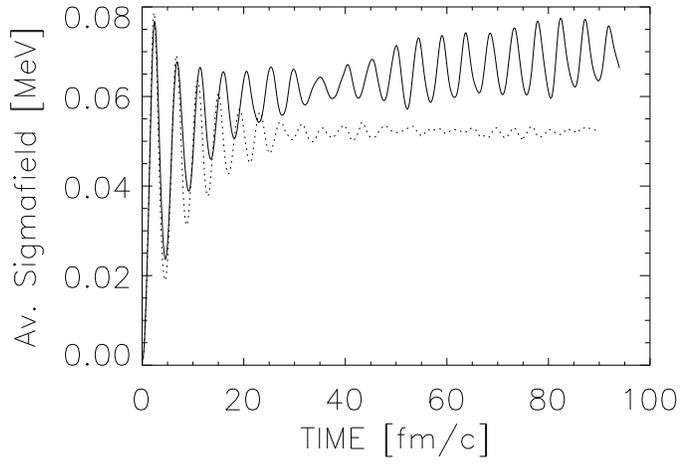,width=10cm}}}
\caption{Additional Figure: $\sigma$-field (averaged over the entire
lattice) as a function of time for $a=0.2$~fm (full line) and $a=1$~fm
(dotted line).}
\label{sigavert}
\end{figure}

Furthermore, we found that the $\sigma$-field relaxes to a value more
closely to $f_\pi$, the correct vacuum value, if $a$ is small. We obtained
$\langle\sigma\rangle=0.80f_\pi$ for
$a=0.2$~fm while $\langle\sigma\rangle=0.56f_\pi$ for $a=1$~fm; both values
determined at $t=120$~fm; see Fig.~\ref{sigavert}.
This observation is opposite to the
expectation in the static, thermal equilibrium case~\cite{Bodeker:1995pp}.
In the dynamical case the
high-frequency classical fluctuations, which would
otherwise mimic finite-temperature contributions to the
potential, cool more efficiently by transfering more of their energy
to the soft modes if $a$ is small (as is obvious from Fig.~\ref{avfield_t80a1}
and~\ref{avfield_t80}). This is not in contradiction to the fact that
for $t\rightarrow\infty$ one should have equipartitioned energy (classically)
as the reverse process, energy transfer from soft to hard modes,
occurs on much longer time-scales.
Thus, we employ $k_c\simeq1.5m_\sigma$
in the simulation of the model on the lattice (described above).
In this way, we account for the full non-linear evolution of those
parametrically amplified modes. 

The modes above $k_c$ could be accounted for by integrating them out of the
partition function, assuming e.g.\ a thermal
distribution~\cite{Boyanovsky:1993pf,Bodeker:1995pp}.
The scattering of modes with
$k\le k_c$ off modes with $k>k_c$ will then introduce fluctuation and
dissipation terms into the classical equations of motion.
Alternatively, the interactions between classical modes and
quantum fluctuations could even be treated explicitly, such that the
latter effectively act as a heat-bath for the former~\cite{Cooper:1997ii}.
That is, however, beyond the scope of the present manuscript.

We now reintroduce the coupling of the $\sigma$-field to the quarks.
For the constituent quark mass $m_q=g\sqrt{\sigma^2+\pi^2}$ in the
vacuum to be one third of the proton mass, we employ $g=3.3$.
Furthermore, dropping the $\pi$-field and again assuming a homogeneous
$\sigma$-field, one arrives at the following equation of motion for modes
of the auxilliary field $\psi$~\cite{Greene:1999nh},
\be \label{parresf}
\ddot\psi_k+\left\{ k^2+g^2\left(\sigma+v\right)^2
-ig\dot\sigma\right\}\psi_k=0\quad,
\ee
where $\sigma=\sigma_0\sin m_\sigma t$, as above.
The quark wavefunction can be obtained from $\psi_k$ as
\bea
q(\vec{x},t) &=& \sum_r\left[i\gamma\cdot\partial+g\sigma\right]\int
\frac{d^3k}{(2\pi)^3 2\Omega_k(t=0)} \nonumber\\
& & \times b_{\vec{k},r}
 e^{i\vec{k}\cdot\vec{x}}\psi_k(t) R_r\quad.
\eea
$R_r$ denote the two eigenvectors of $\gamma^0$ with eigenvalue +1.
Again, the real part of the expression in curly brackets in
eq.~(\ref{parresf}) denotes the oscillator frequency $\Omega_k^2$.
As initial condition we employ
\be
\psi_k(t=0) = {\cal N}_k\quad,\quad \dot\psi_k(t=0) = -i\Omega_k{\cal N}_k
\quad,
\ee
with $1/{\cal N}_k=\sqrt{2\Omega_k\left(\Omega_k+g\sigma+gv\right)}$.
The occupation number of each mode is
\be
n_k = \frac{1}{2}-\frac{k^2}{\Omega_k}{\rm Im}\left(
\psi_k\dot\psi_k^*\right)-g\frac{\sigma+v}{2\Omega_k}\quad,
\ee
and the energy density is given by expression~(\ref{edenseq}), with
a factor 24 (instead of 3) in front of the integral.

\begin{figure}[htp]
\centerline{\hbox{\epsfig{figure=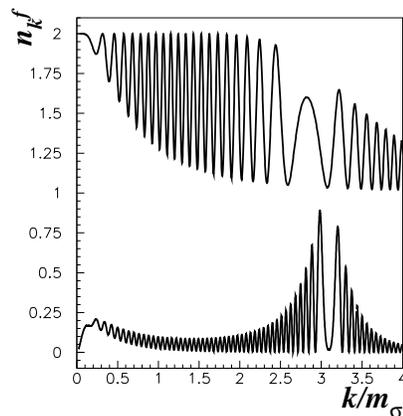,width=6cm}}}
\caption{Occupation numbers of various Fourier modes of the $\psi$-field
after 10 oscillations of the $\sigma$-field with amplitude $\sigma_0=v$
(bottom curve) and $\sigma_0=m_\sigma$ (top curve; $1+n^f_k$ is shown
in that case).}
\label{occno_f}
\end{figure}  
Fig.~\ref{occno_f} shows the occupation number of the various fermion
modes. Due to the Pauli principle
exponential growth of the occupation numbers
is not possible. However, while the incoherent process
$\sigma\rightarrow q\overline{q}$ with ``on-shell'' kinematics is
forbidden (the quark mass, averaged over one oscillation of the $\sigma$-field,
is $m_q=g(f_\pi+2\sigma_0/\pi)>m_\sigma/2$),
the $\sigma$-modes can act as coherent field to produce fermions.
This is because the quark-mass becomes small during some time-interval
of the coherent $\sigma$-oscillations and the frequency $\Omega_k$ of
the quark wavefunction can drop below $m_\sigma$ (one can show analytically
that no quarks are being produced if the $\sigma$-field is static
relative to $\psi_k$).
As opposed to the bosonic occupation numbers, small-$k$ quarks are
produced if the amplitude of the driving $\sigma$-field is large.
However, even in this very simple analysis disrespecting total
energy conservation the long-time energy density of the quarks is
well below that of the pions. Only if the $\sigma$-oscillations were
strongly damped, e.g.\ by a rapidly expanding background, quark-pair
production may be as important as pion production. The
discussion of the condensate relaxation in an expanding
metric is out of the scope of the present manuscript,
see e.g.~\cite{Scavenius:1999zc,Asakawa:1995wk}.

In summary, we have performed numerical simulations of the classical dynamical
evolution of the chiral condensates in strong interaction on a lattice.
In particular, we have focused on the situation where the chiral fields
are rapidly quenched from an initial state with fluctuations.

We confirmed previous results in that the spinodal decomposition
process does not lead to formation of big domains of coherent
$\pi$-field. This remains true also at later times if indeed
only the zero-mode can be described classically, i.e.\ for rather large
lattice spacing $a\ge1$~fm. However, if higher modes are also treated
classically by decreasing the lattice spacing to $\simeq 0.1-0.2$~fm,
long-range correlations (domains) do form as a consequence of
parametric resonance and the nonlinear dynamics of those exponentially
amplified modes.
Despite evident limitations of the effective model that we employed,
we think that the strong amplification
of ``hard'' modes of the condensates, which as a consequence evolve
nonlinearly and eventually drive the amplification of the ``soft''
modes, may occur also in more realistic models.

Finally, we have studied very briefly the effect of coupling the chiral
fields to constituent
quarks. Despite the fact that the Dirac vacuum is not stable
in the presence of coherent classical $\sigma$-field oscillations
with big amplitude or high frequency,
the energy density of the quarks remains well below that of the
$\pi$-field modes at late time.
Thus, the coupling of the condensates to quarks is not likely to 
destroy the ``pseudoscalar laser'' emerging if an initial state
with fluctuations is rapidly ``quenched'' to the spontaneously
broken chiral symmetry.
\acknowledgements
We are happy to thank C.~Greiner, M.~Gyulassy,
A.~Jackson, L.~McLerran, I.~Mishustin, R.~Pisarski, D.~Rischke,
and D.~Son for fruitful discussions and useful comments; and
L.~McLerran for initiating this study.
A.D.\ acknowledges support from the DOE Research Grant, Contract No.\
De-FG-02-93ER-40764.

\end{document}